\documentstyle[12pt,aasms]{article}
\slugcomment{{\em Astrophysical Journal Letters}, 440, L41 (1995).}
\begin{document}
\title{A Supernova at $z = 0.458$
and Implications for Measuring the Cosmological Deceleration}
\author{ 
S. Perlmutter\altaffilmark{1,2},
C. R. Pennypacker\altaffilmark{1,2},
G. Goldhaber\altaffilmark{1,2},
A. Goobar\altaffilmark{2,3},
R. A. Muller\altaffilmark{2},\\
H. J. M. Newberg\altaffilmark{1,2,4}, 
J. Desai\altaffilmark{2},
A. G. Kim\altaffilmark{1}, 
M. Y. Kim\altaffilmark{2},
I. A. Small\altaffilmark{1,2}, 
B. J. Boyle\altaffilmark{5},\\
C. S. Crawford\altaffilmark{5},
R. G. McMahon\altaffilmark{5}, 
P. S. Bunclark\altaffilmark{6},
D. Carter\altaffilmark{6}, 
M. J. Irwin\altaffilmark{6},\\
R. J. Terlevich\altaffilmark{6},
R. S. Ellis\altaffilmark{7},
K. Glazebrook\altaffilmark{7},
W. J. Couch\altaffilmark{8},
J. R. Mould\altaffilmark{9},
T. A. Small\altaffilmark{9},\\
and R. G. Abraham\altaffilmark{10}
}
\altaffiltext{1}{Center for Particle Astrophysics
and Space Sciences Laboratory, U.C. Berkeley, California 94720}
\altaffiltext{2}{Lawrence Berkeley Laboratory, 
Berkeley, California 94720}
\altaffiltext{3}{Swedish Natural Science Research Council (NFR) Fellow}
\altaffiltext{4}{Fermilab, Batavia, Illinois 60510}
\altaffiltext{5}{Institute of Astronomy, Cambridge, United Kingdom}
\altaffiltext{6}{Royal Greenwich Observatory, Cambridge, United Kingdom}
\altaffiltext{7}{University of Durham, Durham, United Kingdom}
\altaffiltext{8}{University of New South Wales, Sydney, Australia}
\altaffiltext{9}{California Institute of Technology, Pasadena, California
91125}
\altaffiltext{10}{Dominion Astrophysical Observatory, Victoria, B.C., Canada}
\begin{abstract}
We have begun a program to discover high-redshift supernovae ($z \approx$
0.25--0.5), and study them with follow-up photometry and spectroscopy.  
We report here our first
discovery, a supernova at $z = 0.458$.  
The photometry for this supernova closely matches the lightcurve 
calculated for this redshift from the template of well-observed 
nearby Type Ia supernovae.  
We discuss the measurement of the deceleration parameter $q_0$ using 
such high-redshift supernovae, and give
the best fit value assuming this one supernova is
a normal, unextincted Type Ia.
We describe the main sources of error in 
such a measurement of $q_0$, and ways to reduce these errors.
\end{abstract}

\vspace*{0.5in}
\section{Introduction}

\vspace{-0.11in}
For over 25 years Type I and Type Ia supernovae (SNe Ia) have been
studied as potential  
``standard candles'' for distance measurements (for a 
review, see Branch \& Tammann 1992).\nocite{Bra:Tam}
Recently, Branch \& Miller (1993)\nocite{Bra:type} and Vaughan {\em et al.}
(1994)\nocite{Vau:Ia} emphasized the narrow distribution of absolute
magnitudes at maximum light for the subset of SNe Ia that have well-measured
lightcurves, and are not unusually red or spectroscopically peculiar. 
This subset of 27 ``normal'' SN Ia has a dispersion,
$\sigma_{M_B} = 0.3$, that is completely accounted for by 
measurement errors, so the {\em intrinsic\/} dispersion should be 
smaller still.  Sandage \& Tammann (1993)\nocite{San:Tam}
estimated $\sigma_{M_B}^{intrinsic} < 0.2$
by simulating the effects of intrinsic dispersion on Malmquist 
bias (using a slightly different set of SNe).

Phillips (1993)\nocite{Phi:Absol}  
noted a correlation between absolute magnitude and 
lightcurve decay time in 9 well-studied SNe Ia (including a few 
``peculiars''). Magnitude corrections based on such a
correlations can ``sharpen up'' the standard candle by calibrating
it, thus improving on the 
current relatively narrow dispersion and also making it possible
to include some peculiar SNe.  For this paper, however, 
we will take the intrinsic dispersion without this calibration to be 
$\sigma_{M_B}^{intrinsic} \approx 0.25$, and we will take the 
error in the mean absolute magnitude for the 27 good SNe Ia to be 
$\sigma_{M_B}^{mean} = \sigma_{M_B}/ \sqrt{27} \approx 0.06$ (note
that these are somewhat more conservative estimates that those of
Sandage \& Tammann (1993)).

\vspace{-.03in}
Type Ia's are on average the brightest SNe
and therefore could be used to 
measure large cosmological distances and, in particular,
the deceleration parameter $q_0$ of the 
expanding universe (Tammann 1979;\nocite{Tam:Astro} 
Colgate 1979)\nocite{Col:Super}.
With many observable 
features in their spectra and their lightcurves, SNe Ia have 
an advantage as
standard or calibrated candles because they can be checked one-by-one for 
evolutionary differences at high redshift.  For example, the
lightcurve decay time or the blueshift of spectral features may indicate
explosion strength, and can be checked for systematic differences from
nearby supernovae.

\vspace{.01in}
To use SNe as tools for cosmology, we have developed a 
strategy to 
find and study high-redshift SNe systematically, using new 
wide-field, high-resolution CCD cameras and image analysis techniques that 
provide the rapid response time necessary to follow these short-lived events 
(Perlmutter {\em et al.} 1993).\nocite{Perl:Syst}
In 1992 March and April, we demonstrated this strategy using 
the Isaac Newton 2.5 meter Telescope (INT) on La Palma, and found a 
SN at $z = 0.458$.  The farthest previously detected SN was found
at $z = 0.31$ in the SN search of Norgaard-Nielsen {\em et al.} 
(1989).\nocite{Nor}

\vspace{-0.19in}
\section{Observations and Data Reduction}

\vspace{-0.27in}
During 1992 March 24 -- 28, we observed 54 high-galactic latitude
fields, 43 of them centered on
high-redshift clusters, using 10-minute 
$R$ band exposures in
seeing better than 2 arcsec. 
Approximately 200 galaxies were visible 
in the redshift range $z =$ 0.25 -- 0.5 in each
10.5 arcmin $\times$ 11.5 arcmin image to a limiting magnitude
of $R \approx 23$ ($4 \sigma$ detection).  
During 1992 April 21 -- May 2 the same fields were re-observed,
in pairs of 7-minute exposures offset 
by $\sim$6 arcsec to facilitate rejection of cosmic rays and CCD defects.
Following each night's observations
we used semi-automated image analysis software to search
for new point-sources that had appeared since 
the reference images were taken.

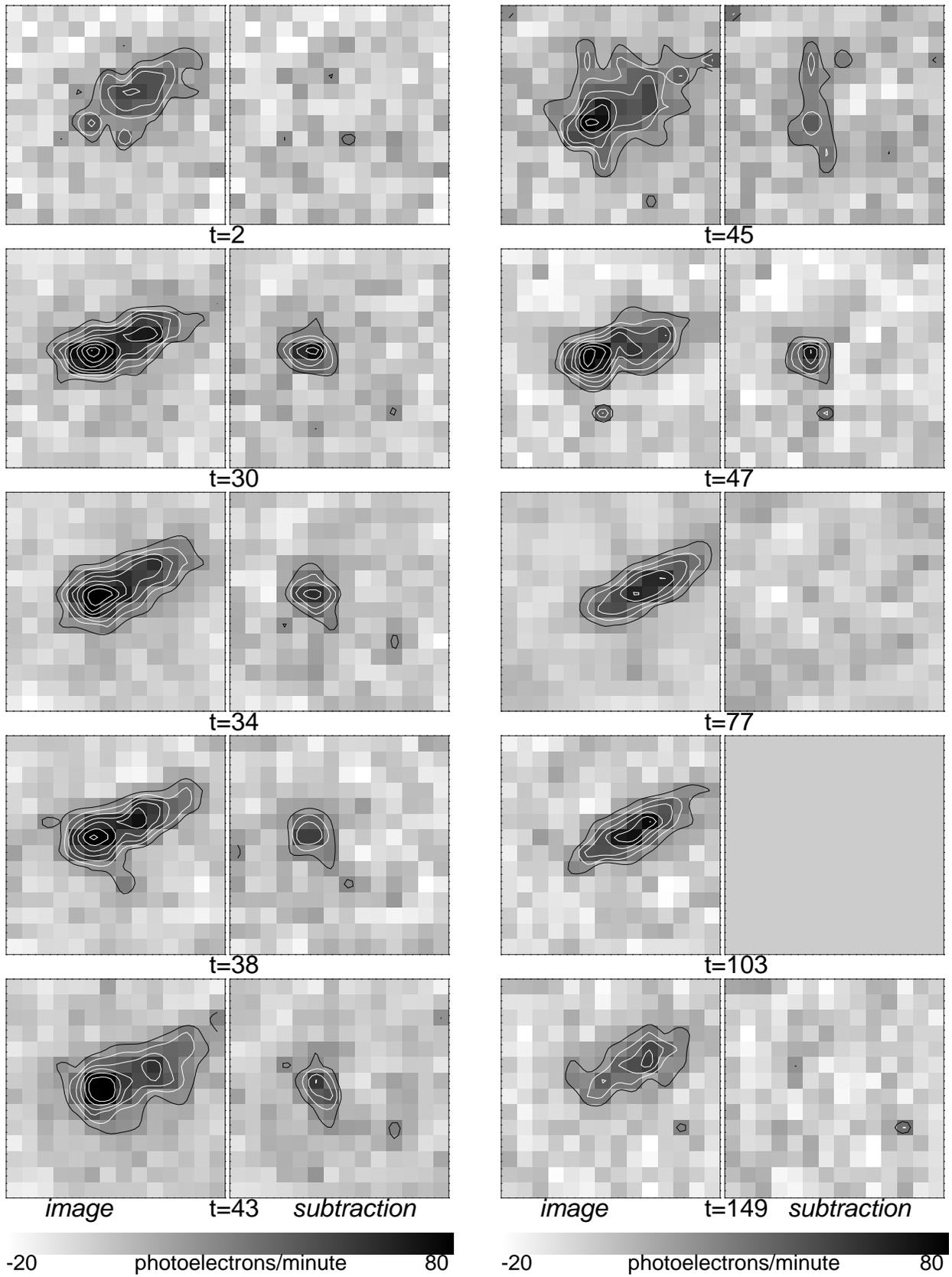
\begin{figure*}
%[tb]
% \psfig{figure=contours.ps,height=4.2in}
%\plotone{[astro.saul]contours.ps}
\caption[figurecaption]{}
%{(a) Contour plots of host galaxy with and without
%supernova. The time, $t$, is JD$\mbox{}-2448707$.  The 
%contours start at 22 photoelectrons/minute/pixel, with each additional
%contour representing 11 photoelectrons/minute/pixel.  
%Note that these images are normalized
%to the same sky transmission but not matched in seeing.
%(b) The same images after subtracting off the host galaxy reference image.}
\label{contour}
\end{figure*}

\vspace{0.04in}
With good seeing and 0.57 arcsec pixels, we were able to 
distinguish candidate SNe from other variable sources (e.g. QSO's) 
by looking for a resolved host galaxy.  The time between 
observations at the INT and the completion of analysis in Berkeley was 
typically less than 2 days, so interesting candidates could trigger 
follow-up observations.  The best of these SN candidates (SN 1992bi)
was found at R.A. = $16^h8^m28\fs4$,
Dec. = $+39\deg54\arcmin58\arcsec$ (equinox 1950.0), 
$1\farcs5$ east and $0\farcs5$ north from the 
core of an $R \approx 21.3$ magnitude galaxy 
(Pennypacker, {\em et al.} 1992).\nocite{Pen:in}  
We observed the candidate on 5 more nights 
during the 2.5 weeks after discovery and on 4 nights 
during the 9 months after it had faded below detectability.  
Figure~\ref{contour} (Plate L\#)
shows the images of the host 
galaxy before, during, and after the event.  To the right of each image
is the same 
image after subtracting off a ``reference'' image from Day 103 
of the host galaxy alone.

\vspace{0.04in}
The subtraction images of 
Figure~\ref{contour} were used to measure the photon flux $f_R^{\rm 92bi}$
of the SN candidate on each observation.  
The Day 396 flux is taken as the best estimate of the host galaxy light
and is subtracted out of each $f_R^{\rm 92bi}$ measurement.
The night-to-night ``relative'' 
measurement error is approximately 9\% at maximum light, and is 
primarily due to photon noise of 
the background sky.
A smaller error contribution, 
$\sim$5\% at maximum light,
is due to slight mismatches of the two images before subtraction of the 
host galaxy reference image from the measured image of the galaxy 
plus SN.  These errors were determined by extensive
testing of the plausible range 
%\pagebreak
of matching parameters (position offset, seeing, and transmission ratio) for 
their effect on the resulting magnitude measurements.
To obtain the absolute measurement error we must add 
(in quadrature with these relative errors) an overall error
due to the sky noise of the Day 396 reference image, amounting to
$\sim$13\% near peak.
Table~\ref{magtable} lists for each observation
the flux $f_R^{\rm 92bi}$ and 
relative error of the SN candidate measured
from the subtraction images (note that $f_R^{\rm 92bi}$ can be negative due to 
photon noise of the subtracted host galaxy and subtracted sky).

The calibration to $R$ magnitude
was made on day 149 using the INT/EEV5 camera 
to image M92 and the SN field consecutively 
(repeated 3 times to check photometric
stability) with the same airmass within 0.02.  
These observations were repeated in the $V$ and $I$ bands.
Our relative photometry agreed with 
the calibration of 23 stars in M92 (Christian {\em et al.} 1985,\nocite{Chr} 
revised L. Davis, private communication) with
a root-mean-square error of 0.005 mag, and yielded the following
calibration:  $m_R^{\rm 92bi} = -2.5 \log f_R^{\rm 92bi} + 28.20$, 
where $f_R^{\rm 92bi}$ is measured
in units normalized to the photoelectrons/minute observed
at the INT 2.5-meter on Day 103.  
The uncertainty in transferring this standard to each
individual night's observation is $\sigma_{cal} = 0.02$ mag, primarily due
to the photon noise on $\sim$20 bright stars in the SN field
found in both the standard image and each night's images.
(Note that the first order extinction is calibrated out in this procedure,
and that the extinction color dependence is negligible, because
the calibration observations were in the $R$ band and at an airmass within
0.2 of each of the SN observations.  The color term is $<0.01$ implying that 
our instrumental $r$ band is quite close to the calibration's Cousins $R$.)

Twelve attempts to obtain the SN spectrum at four observatories 
around the world were all unsuccessful due to weather (and one instrument 
failure).  On 1992 August 29, we obtained two 1800-second spectra of the 
host galaxy using the
Low Dispersion Survey Spectrograph 2 on the 
William Herschel 4.2-meter telescope, and 
determined the redshift $z = 0.458 \pm 0.001$.  
Both spectra show a strong emission line at 5437 Angstroms.
Identifying this line as [OII]$\lambda 3727$ at $z=0.458$ allows 
the weak absorption features longwards of it
to be identified with Ca H, Ca K, and ${\rm H}\delta$ (see 
Figure~\ref{spectrum}).

\vspace{-0.2in}
\section{Analysis}

\vspace{-0.27in}
Traditionally one compares photometry of 
high-redshift objects to photometry of nearby objects by calculating a 
$K$-correction to account for the different parts of the spectrum that fall in 
one given filter band for objects at different redshifts.  
A more robust procedure in this case is to compare the photometry of 
Table~\ref{magtable} to the apparent $R$-band lightcurve, $m_R(t)$, 
calculated for $z=0.458$ from the standard lightcurve of well-observed 
nearby SN Ia {\em in the B-band}.   This standard lightcurve,
$M_B(t)$, is an
apparent magnitude versus 
redshift measurement and thus implicitly depends on the Hubble constant 
chosen.  However, the expression [$M_B(t) - 5 \log H_0$] that will appear
bracketed together throughout this Letter is independent of the choice
of Hubble constant, as is the resulting measurement of $q_0$.
Since events seen at $z=0.458$ last 1.458 times as long as they do locally,
this lightcurve term must be time dilated to [$M_B(t/(1+z)) - 5 \log H_0$].
\nopagebreak
The calculated distant $R$-band lightcurve (for $\Lambda = 0$) is thus:
\pagebreak
\begin{eqnarray*}
	m_R(t) & = & [M_B(t/(1+z)) - 5 \log H_0] 
+ \Delta m_{RB} + A_R + 25
  \\
& &  \mbox{} + 5 \log {
\left(
{c}{ q_0^{-2}}
\left[
1 - q_0 + q_0z + (q_0 - 1) ( 2 q_0 z + 1 )^{\frac{1}{2}} 
\right]
\right),
}
\label{mR}
\end{eqnarray*}
where $A_R \simeq 0.006$ is the $R$ extinction in our 
Galaxy at $(l,b) = (63.27,47.24)$ (Burstein 
\& Heiles 1982;\nocite{Bur} Burstein, private 
communication).  The $\Delta m_{RB}$ term is the analog of the traditional 
$K$-correction:
\begin{eqnarray*}
\Delta m_{RB} &  =  & -2.5 \log
    \left(
    \frac
       {\int F_\lambda^{A0V}(\lambda)S_B(\lambda)d\lambda}
       {\int F_\lambda^{A0V}(\lambda)S_R(\lambda)d\lambda}
\;\;    
    \frac
       { \int F_\lambda^{SN}(\lambda/(1+z))S_R(\lambda)d\lambda}
	{\int F_\lambda^{SN}(\lambda)S_B(\lambda)d\lambda}
    \right)
+ 2.5 \log (1+z)
\\
  &  =  & -2.5 \log
    \left(
    \frac
       {\int F_\lambda^{A0V}(\lambda)S_B(\lambda)d\lambda}
       {\int F_\lambda^{A0V}(\lambda)S_R(\lambda)d\lambda}
\;\;    
    \frac
       { \int F_\lambda^{SN}(\lambda')S_R(\lambda' (1+z))d\lambda'}
	{\int F_\lambda^{SN}(\lambda)S_B(\lambda)d\lambda}
    \right),
\label{dRB}
\end{eqnarray*}
where $S_B(\lambda)$ and $S_R(\lambda)$ are the 
response functions in the $B$ and $R$ bands,
and $F_\lambda^{SN}(\lambda)$ and $F_\lambda^{A0V}(\lambda)$ are 
the flux per unit
of wavelength of the supernova and the standard AOV star SED for 
which $B = R = 0$ in the Landolt system.

The correction 
$\Delta m_{RB}$ accounts for the different zero points of the $R$ and $B$
magnitude systems, and for the the redshifting of the SN spectrum.
We calculate $\Delta m_{RB} = -0.7 \pm 0.05$ using numerical integration with
spectra of 3 SNe Ia with
overlapping epochs from
15 days before to 15 days past maximum (in the SN rest frame).
The error represents the
variation for a given SN over time; the SN-to-SN 
variation at maximum is 0.02 mag.  
The relatively small variation of $\Delta m_{RB}$ over time is due
to the approximate match at $z=0.458$ between the
observer's $R$ response function and the SN rest-frame's $B$ response 
function, i.e.
$S_R(\lambda (1+z)) \approx 
S_B(\lambda)$.
At $z=0.458$, $\Delta m_{RB}$ is therefore less sensitive to the
precise knowledge of the changing SN spectrum, an 
advantage over the usual $K$-correction.
(The calculation of $\Delta m_{RB}$
was checked by reproducing the $K$-corrections of Hamuy {\em et al.} (1993) 
\nocite{Ham:kcorr} to within 0.001 mag for the 3 SNe.)

Taking [$M_B(t_{max}) - 5\log (H_0/75)] = -18.86 \pm 0.06$, where the error 
is the error in the mean of the distribution as discussed above
(Branch and Miller 1993, Vaughan {\em et al.} 1994), and
the template standard $B$ lightcurve, $M_B(t) - M_B(t_{max})$, of
\nocite{Lei:super}Leibundgut (1991 and private communication), we calculate
the lightcurves shown in
Figure~\ref{lightcurve} for two values of the
deceleration parameter, 0 and 0.5.  At $z=0.458$ 
a standard candle is 0.25 mag fainter (for $\Lambda = 0$)
in an empty universe, $q_0 = 0$, than in a critically closed universe,
$q_0 = 0.5$.

Figure~\ref{lightcurve} also shows the measured 
photometry data from Table~\ref{magtable}.  When
we perform a $\chi^2$ fit of the  
the data points of Table~\ref{magtable} to the calculated lightcurve,
$f_R(t)$, varying the values of $q_0$ and 
the time offset, $\Delta t$, we find the best
fit at $q_0=0.07$.
If there is significant
host galaxy extinction then $q_0$ would be larger.
The individual error bars shown in Figure~\ref{lightcurve} are the
relative errors used in the fit, while the errors from the 
reference image photometry and calibration contribute to the overall
error bar $\sigma_c$ (discussed below).
Note that the pre- and post-SN measurements
constrain the time-offset fit.

Although the photometry data shown in Figure~\ref{lightcurve} 
are clearly consistent with the 
lightcurve of an SN Ia
at $z = 0.458$, several alternative identifications were considered.  
It is very unlikely (Prob $<$ 1\%) that this event is a projected 
foreground variable star in our Galaxy or a projected background quasar or 
AGN.  The lack of repeated flare events makes these identifications still 
more unlikely.  The data {\em are} also consistent with an 
unusually bright SN IIL.
Assuming the currently favored cosmologies with $q_0 < 2$ (or, equivalently,
$H_0 > 40 \; {\rm km}\;{\rm sec}^{-1}\;{\rm Mpc}^{-1}$ and 
$\tau_{\rm universe} > 12$ Gyr; see Carroll, Press, \& Turner 1992), 
we can estimate the relative probability of a SN IIL identification as follows:
First, we note that the event's absolute magnitude could not be
more than 0.7 magnitudes fainter than the SN Ia value
$M_{B}(t_{max})$ used above, because $q_0$ would then have to be 
$>2$ to fit our photometry data.  We then compare, in nearby
spiral galaxies, the number of
SNe Ia and the number of SNe IIL brighter than this minimum:
$M_{min} \equiv M_{B}(t_{max}) +0.7 + 0.3$ (the additional 0.3
magnitude allows for measurement error).  The result is
$\sim$10 SNe Ia for every 
SN IIL (see Miller \& Branch 1990), and we use this ratio
as the relative probability of these identifications. 
This is, of course, a rough approximation that assumes that the
SN detection efficiency falls off towards fainter absolute magnitudes
at least as fast at $z=0.458$ as in nearby searches.
Similarly, we estimate $<1/30$ for the 
relative probability that this event is a 
peculiar SN Ia or one of the SNe Ia showing clear evidence of extinction, since
over 30 times as many 
normal, unextinguished SNe Ia brighter
than $M_{min}$ are found in nearby galaxies
(Branch \& Miller 1993).
The identification of this event as a normal SN Ia 
at $z=0.458$ is therefore made with
$\sim$90\% confidence level (for any $q_0 <2$) for the 
purposes of this paper.  The confidence level would be 
greater (and the {\em assumption} of $q_0 <2$ unnecessary)
if spectra or color photometry were available.

The uncertainty in this measurement of $q_0$ has two major 
contributory sources.  The first source is the 
uncertainty, $\sigma_R \approx 0.14$, in the
apparent $R$ magnitude at 
maximum light of the $z=0.458$ SN.
This includes the photon noise and image-matching 
uncertainty in the images observed near 
maximum light, $\sigma_{peak} \approx 0.06$,
and the photon noise and calibration uncertainty
in the Day 396 reference image used to subtract off 
the host galaxy light, $\sigma_{re\!f} \approx 0.13$.  (A better 
reference image will reduce this error source significantly.)

The second and much 
larger contributory source of measurement error enters in when this distant 
photometry is compared with the nearby apparent magnitude-Hubble 
velocity relation, i.e. the nearby measurement of [$M_B - 5 \log 
H_0$].  The uncertainty in the mean of this quantity may be as 
small as $\sigma_{M_B}^{mean} \approx 0.06$, but since we do not 
know where in the $M_B$ distribution our particular SN at 
$z=0.458$ lies, we must add in quadrature an error contribution 
for the intrinsic dispersion $\sigma_{M_B}^{intrinsic} \approx 
0.25$.  {\em This is currently the dominant source of error.}  
Since it is plausible that this is an overestimate of the 
intrinsic dispersion, as discussed above, we will separate out this 
error in the subsequent analysis, and define a $\sigma_o$ to include
only the other sources of error.
Finally, we also include in quadrature an error of $\sigma_{RB} 
\approx 0.05$ for the uncertainty in $\Delta m_{RB}$, so $\sigma_o^2
= \sigma_R^2 + {\sigma_{M_B}^{mean}}^2 + \sigma_{RB}^2$.

\vspace{-0.05in}
The total error in the distant SN 
apparent magnitude and in the
comparison with the nearby apparent magnitude-velocity relation 
is $\sigma_{mag} = \pm \sigma_o \pm \sigma_{M_B}^{intrinsic} =
\pm 0.16 \pm 0.25$.  (Figure~\ref{lightcurve}'s $\sigma_c$ is the equivalent
total error in flux.) 
This propagates through to an uncertainty 
on the $q_0$ measurement of $\sigma_{q_0} \approx \sigma_{mag} / 
z$, yielding a measurement of $q_0 = 0.1 \pm 0.3 \pm 0.55$ 
(for $\Lambda = 0$).  This
measurement is stricter in its lower limit than its upper 
limit, since host galaxy extinction would increase $q_0$.  We emphasize
again the assumption of a normal SN Ia.

\vspace{-0.26in}
\section{Discussion}

\vspace{-0.33in}
We wish to stress two main results:

\vspace{-0.09in}
(1)  From a single SN at $z=0.458$ known to be a ``normal'' type Ia,
we could in principle measure $q_0$ with 
an uncertainty comparable to the previous measurements in the 
literature that required larger numbers of objects (galaxies or clusters) 
and needed significant corrections for evolutionary effects (for 
reviews, see Rowan-Robinson 1985\nocite{Row:Cos} 
and Sandage 1988).\nocite{San:Obs}  Clearly the 
measurement of $q_0$ cannot rest on a single distant SN Ia, and 
in this particular case we do not have the color photometry or 
the spectra that would enable us to screen for host galaxy extinction 
and peculiar SNe.  We are continuing to search for distant 
SNe Ia, and we are scaling up the detector size and sensitivity so 
that we can find many high-redshift SNe Ia, and follow their 
lightcurves, colors and spectra over maximum light.  
This will make it possible to compare the 
luminosity function distribution of distant and nearby SNe,
and to identify and reject extinguished SNe.

\vspace{-0.02in}
(2)  The measurement errors that contribute to 
the uncertainty in $q_0$ can all be reduced.  The dominant error 
is due to both our uncertainty in the true width of the SN Ia 
absolute magnitude distribution and our uncertainty in where a 
given distant SN Ia falls in this distribution.  As a few more SN 
Ia past Coma are discovered and studied, we may find that the 
distribution width is smaller than can be seen in the nearby SNe 
where peculiar velocities can mask the Hubble expansion. Hamuy 
{\em et al.} (1994) are working on such a search, 
and the Berkeley Automated Supernova Search
(Muller {\em et al.}, 1992), \nocite{Mul:high}now being moved to 
\nopagebreak
a good astronomical site, may also find and study these SNe.

\vspace{-0.02in}
Alternatively, if the dispersion is in fact 
\pagebreak
$\sigma_{M_B}^{intrinsic} =0.25$ then discovering $N$ more distant 
SNe Ia will make it possible to statistically reduce this error by 
$\sim$$\sqrt{N}$.  The main source of concern here is that the 
distant $M_B$ distribution should be the same as the nearby 
distribution for the comparison to be valid;  it will be 
important to ensure that the distant SN detection limit be 
significantly fainter than the peak of the $M_B$ distribution at 
the redshifts searched to avoid Malmquist-bias distortion of the 
shape of the distribution, particularly if there are rare superluminous
Type Ia's (see the discussion of the possibly superluminous SN 1991T in 
Ford {\em et al.}\nocite{Ford} 1993).

The uncertainty in the $K$-correction (in this case, $\Delta 
m_{RB}$) is the only remaining source of error that cannot be 
improved simply by using longer exposure times, larger 
telescopes, or darker sites.
It therefore may be useful
to measure the lightcurves for future 
nearby SNe Ia using filter sets designed to give
``blue-shifted $R$'' bands for a sampling
of values of $z$. Together with SNe Ia spectra to interpolate between
these $z$ values, these ``blue-shifted $R$'' lightcurves should 
allow accurate determinations of the $K$-corrections.

It will be necessary to check the effects of metallicity and other 
evolutionary changes on the SN Ia absolute magnitude to validate 
measurements of $q_0$ derived from SNe Ia.  
Deep spectra of distant SNe will be necessary to
look for any 
evolutionary changes with respect to spectra of nearby SNe.  
Metallicity effects can be studied by comparing nearby SNe Ia 
found in different metallicity environments.  So far, however, no 
indications of magnitude evolution have been found observationally,
and the standard model of SNe Ia gives theoretical reasons to 
believe that SN Ia magnitudes should not evolve.  
Thus, the prospects for 
obtaining a precise value for $q_0$ from SNe Ia appear good.

[Note added in press:  This project has by now found 6 additional SNe
around $z \sim 0.4$ (e.g., Perlmutter {\em et al} 1994).  
With these it should be possible to use 
lightcurve decay-time calibration or lightcurve
shape calibration, recently shown to yield 
$\sigma_{M_B}^{intrinsic}$ as small as
0.1 (Hamuy {\em et al.} 1994) to 0.2
(Riess {\em et al.} 1994).]

%\acknowledgments

M. Bessell,
B. Leibundgut, B. Schmidt, M. Hamuy, and D. Branch
kindly supplied unpublished data, spectra, programs, and some results 
for the calculation of $\Delta m_{RB}$.  We thank 
D.Burstein, K.Chambers, R.Clegg, 
M.De Robertis, S.Kulkarni, P.Meikle, F.Sanchez, and P.Vilchez for
observational efforts.  
This work has support from the U.S. Dept. of Energy (DE-AC03-76SF000098)
and National Science Foundation (ADT-88909616).

\begin{table}
\begin{minipage}[t]{6in}
\begin{tabular}{c c r c c} \tableline
Day & Telescope &
Exposure & $f_R^{\rm 92bi}$ \\ 
JD-2448707 & /Camera &
(seconds) & (relative flux)\tablenotemark{\em a} \\
\tableline \tableline
0& INT/EEV5 & 600 & $-18 \pm 42$ \\ 
2& INT/EEV5 & 600 & $\mbox{  } 25 \pm 25$ \\ 
30& INT/EEV5 & 2$\times$420 &$ 260 \pm 21$ \\ 
34& INT/EEV5 & 2$\times$420 & $251 \pm 22$ \\
38& INT/EEV5 & 2$\times$480 & $241 \pm 29$ \\
43& INT/EEV8 & 2$\times$300 & $213 \pm 23$ \\ 
45& INT/EEV8 & 300 & $169 \pm 41$ \\
47 & INT/EEV8 & 300 & $215 \pm 34$ \\
77& Palomar/{\sc Cosmic} & 300 & $\mbox{  } 28 \pm 44$  \\
103 & INT/EEV5 & 5$\times$$\sim$170 &$\mbox{  } 20 \pm 19$ \\
148 & INT/EEV5 & 6$\times$300 & $\mbox{  } 12 \pm 22$ \\
330 & INT/Ford & 6$\times$320 & $-34 \pm 22$ \\ 
396 & INT/EEV5 & 7$\times$400 & ... \\ 
\tableline
\end{tabular}
\end{minipage}
\tablenotetext{a}{The host galaxy flux from the image of Day 396
has been subtracted off.  Relative flux units
are normalized to the photoelectrons/minute observed
at the INT 2.5-meter on Day 103.  The calibration
is: $m_R^{\rm 92bi} = -2.5 \log f_R^{\rm 92bi} + (28.20 \pm 0.02)$.
The listed relative error includes 
sky photon noise and image
matching error for each individual image.  
An overall error of 30 units should
be added in quadrature to account for the
sky noise of the reference image. }
 \caption{Observation log and supernova flux}\label{magtable} 
\end{table}

\newpage
\begin{center}
{\bf FIGURE CAPTIONS}
\end{center}
\medskip

\noindent {\bf Figure 1 (Plate L\#):} The left panel of each pair shows
the image of the host galaxy with or without the
supernova. The time, $t$, is JD$\mbox{}-2448707$.  The superposed
contours start at 22 photoelectrons/minute/pixel, with each additional
contour representing 11 photoelectrons/minute/pixel.
Note that these images are normalized
to the same sky transmission but not matched in seeing.
The right panel of each pair shows the
same images after subtracting off the host galaxy image of day 103.
(Days 0, 330, and 396 are not shown.)

\bigskip
\noindent{\bf Figure 2:} Spectrum of the host galaxy, observed on 
1992 August 29 (Day 156) on the William Herschel 4-meter telescope.

\bigskip
\noindent{\bf Figure 3:} Solid curve shows the 
calculated lightcurve, $f_R(t)$,
for the best fit $q_0 = 0.1$, based on the
template $B$ lightcurve for nearby supernovae.  
The dotted curves are $f_R(t)$ for $q_0 = 0.5$ 
(upper curve) and for $q_0 = 0$ (lower curve).  
The photometry points are $f_R^{\rm 92bi}$ from Table~\ref{magtable}
(days 330 and 396 not shown).
The inner error bar, $\sigma_{c}$, shows the combined uncertainty at
maximum light in $f_R(t)$ and in the 
reference image's photometry and magnitude calibration;
the outer error bar includes the intrinsic dispersion, 
$\sigma_{M_B}^{intrinsic}$.  The inset scale shows $R$ magnitudes.
For the best fit $q_0$, the 
peak magnitude is $m_R = 22.2$ on Day 34.
\begin{figure*}
%[tb]
% \psfig{figure=spectrum.ps,height=4.2in}
%\plotone{[astro.saul]spectrum.ps}
 \caption[figurecaption]{}
%{Spectrum of the host galaxy, observed on 
%1992 August 29 (Day 156) on the William Herschel 4-meter telescope.}
 \label{spectrum}
\end{figure*}
\begin{figure*}
%[tb]
%\psfig{figure=candcurve.ps,height=4.0in}
%\plotone{[astro.saul]candcurve.ps}
 \caption[figurecaption]{}
%{Solid curve shows the 
%calculated lightcurve, $m_R(t)$,
%for the best fit $q_0 = 0.1$, based on the
%template $B$ lightcurve for nearby supernovae.  
%The dotted curves are $m_R(t)$ for $q_0 = 0.5$ 
%(upper curve) and for $q_0 = 0$ (lower curve).  
%The photometry points are from Table~\ref{magtable}.
%The inner error bar, $\sigma_{c}$, shows the combined uncertainty at
%maximum light in $m_R(t)$ and in the 
%overall magnitude calibration of the photometry points;
%the outer error bar includes the intrinsic dispersion, 
%$\sigma_{M_B}^{intrinsic}$.  For the best fit $q_0$, the 
%peak magnitude is $m_R = 22.2$ on Day 34.
%}
 \label{lightcurve}
\end{figure*}
\end{document}